\documentclass[pdftex,cite,aps,prl,superscriptaddress,amsmath,amsfonts,amssymb,twocolumn,showpacs]{revtex4}
\usepackage{wasysym}
\usepackage{graphics}
\usepackage{subfigure}
\usepackage[pdftex]{graphicx}
\begin{document}
\title{Tailoring optical nonlinearities via the Purcell effect} 
\author{Peter Bermel, Alejandro Rodriguez, John D.
  Joannopoulos, Marin Solja\v{c}i\'{c}} 
\affiliation{Department of Physics, Massachusetts Institute of
  Technology, Cambridge, Massachusetts, 02139, USA}
\date{\today}

\begin{abstract}
  We predict that the effective nonlinear optical susceptibility can
  be tailored using the Purcell effect.  While this is a general
  physical principle that applies to a wide variety of nonlinearities,
  we specifically investigate the Kerr nonlinearity.  We show
  theoretically that using the Purcell effect for frequencies close to
  an atomic resonance can substantially influence the resultant Kerr
  nonlinearity for light of all (even highly detuned) frequencies.
  For example, in realistic physical systems, enhancement of the Kerr
  coefficient by one to two orders of magnitude could be achieved.
\end{abstract}


\maketitle

  Optical nonlinearities have fascinated physicists for many
  decades because of the variety of intriguing phenomena that they
  display, such as frequency mixing, supercontinuum generation, and
  optical solitons~\cite{Drazin:89,Boyd:92}.  Moreover, they enable
  numerous important applications such as higher-harmonic generation
  and optical signal processing~\cite{Boyd:92,Nielsen:00,Agrawal:01}.
  On a different note, the Purcell effect has given rise to an entire
  field based on studying how complex dielectric environments can
  strongly enhance or suppress spontaneous emission from a dipole
  source~\cite{Purcell:46,Kleppner:81,Ryu:03,Bermel:04,Englund:05}.
  In this Letter, we demonstrate that the Purcell effect can also be
  used to tailor the effective nonlinear optical susceptibility.
  While this is a general physical principle that applies to a wide
  variety of nonlinearities, we specifically investigate the Kerr
  nonlinearity, in which the refractive index is shifted by an amount
  proportional to intensity.  This effect occurs in most materials,
  modeled here as originating from the presence of a collection of
  two-level systems.
  We show theoretically that using the Purcell effect for frequencies
  close to an atomic resonance can substantially influence the
  resultant Kerr nonlinearity for light of all (even highly detuned)
  frequencies.

In hindsight, the modification of nonlinearities through the Purcell
effect could be expected intuitively: optical nonlinearities are
caused by atomic resonances, hence varying their strengths should
influence the strengths of nonlinearities as well.  Nevertheless, to
the best of our knowledge, this interesting phenomenon has not thus
far been described in the literature.  Moreover, as we show below, it
displays some unexpected properties.  For example, while increasing
spontaneous emission strengthens the resonance by enhancing the
interaction with the optical field, it actually makes the optical
nonlinearity weaker.  Furthermore, phase damping (e.g., through
elastic scattering of phonons), which is detrimental to most optical
processes, plays an essential role in this scheme, because in its
absence, these effects disappear.

A simple, generic model displaying Kerr nonlinearity is a two-level
system.  Its susceptibility has been calculated to all orders in both
perturbative and steady state limits~\cite{Boyd:92}.  However, this
derivation is based on a phenomenological model of decay observed in a
homogeneous medium, and does not necessarily apply to systems in which
the density of states is strongly modified, such as a cavity or a
photonic crystal bandgap.  Following an approach similar to
Ref.~\onlinecite{JohnQu:96}, the validity of this expression can be
established from a more fundamental point of view.  Start by
considering a collection of $N$ two-level systems per unit volume in a
photonic crystal cavity, whose levels are labeled $a$ and $b$.  The
corresponding Hamiltonian is given by the sum of the self-energy and
interaction terms ($H_0$ and $V(t)$, respectively).  Using the
electric dipole approximation, one obtains:
\begin{equation}
\label{myhamiltonian}
H= H_0 + V(t) = \hbar\left[\omega_a \sigma_{aa} + \omega_b \sigma_{bb} + \Omega(t) \sigma_{ab} + \Omega^*(t)\sigma_{ba}\right],
\end{equation}
where $\sigma_{ij}=c_i^{\dagger}c_j$ is the operator that transforms
the fermionic state $j$ to the fermionic state $i$,
$\Omega(t)=-\vec{\mu}\cdot \vec{E}(t)/\hbar$ is the Rabi amplitude of
the applied field as a function of time, and the scalar dipole
moment $\mu$ is defined in terms of its projection along the applied
field $\vec{E}(t)$.  In general, if this system is weakly coupled to
the environmental degrees of freedom, then the timescale for the
observable dynamics of the system is less than the timescale of the
``memory'' of the environment.  In this case, information sent into
the environment is irretrievably lost -- this is known as the
Markovian approximation~\cite{Preskill:04}.  The dynamics of this
system can then be modeled by the Lindbladian ${\cal L}$, which is a
superoperator defined by $\dot{\rho}\equiv{\cal L}\left[\rho\right]$.
In general, one obtains the following master equation from the
Lindbladian:
\begin{equation}
\dot{\rho}=-(i/\hbar)\left[H,\rho\right]+\sum_{\mu} \left[
  L_{\mu} \rho L_{\mu}^{\dagger}-\frac{1}{2}L_{\mu}^{\dagger} L_{\mu}
  \rho - \frac{1}{2} \rho L_{\mu}^{\dagger} L_{\mu} \right]
\end{equation}
Using the only two quantum jump operators that are allowed in this
system on physical grounds -- $L_1\equiv\sigma_{ab}/\sqrt{T_1}$ and
$L_2\equiv\sigma_{bb}\sqrt{\gamma_{\mbox{\tiny
      phase}}}$~\cite{Preskill:04} -- one can obtain the following
dynamical equations:
\begin{equation}
\label{rhobadt}
\frac{d\rho_{ba}}{dt} = -(i\omega_{ba}+T_2^{-1}) \rho_{ba}+i \Omega(t)
(\rho_{bb}-\rho_{aa})
\end{equation}
\begin{equation}
\label{rhoaadt}
\frac{d(\rho_{bb}-\rho_{aa})}{dt} =
-\frac{(\rho_{bb}-\rho_{aa})+1}{T_1}
 - 2i \left[\Omega(t)\rho_{ab}-\Omega^*(t)\rho_{ba}\right],
\end{equation}
where $\omega_{ba}\equiv \omega_b-\omega_a$, $T_1$ is the rate of
population loss of the upper level, and $T_2 = (1/2) T_1^{-1} +
\gamma_{\mbox{\tiny phase}}$ is the rate of polarization loss for the
off-diagonal matrix elements.  The prediction of exponential decay via
spontaneous emission is known as the Wigner-Weisskopf
approximation~\cite{CohenTannoudji:77}.  Although it has been shown
that the atomic population can display unusual oscillatory behavior in
the immediate vicinity of the photonic band
edge~\cite{John:94,Lambropoulos:00}, theoretical~\cite{JohnQu:96} and
experimental considerations~\cite{Bayer:01,Lodahl:04} show that this
approximation is fine for resonant frequencies well inside the
photonic bandgap.  In the rest of this manuscript, this is assumed to
be the case.  Next, one can make the rotating wave approximation for
Eqs. (\ref{rhobadt}) and (\ref{rhoaadt}), and then solve for the
steady state.  If the polarization is defined by $P=N\mu
(\rho_{ba}+\rho_{ab})=\chi E$, where $\chi$ is the total
susceptibility to all orders, one obtains the following well-known
expression for the susceptibility~\cite{Boyd:92,JohnQu:96}:
\begin{equation}
\label{totalchi}
\chi=-\frac{N \mu^2
  (\omega-\omega_{ba}-i/T_2) T_2^2/\hbar}{1+(\omega-\omega_{ba})^2
  T_2^2 + (4/\hbar^2)\mu^2 |E|^2 T_1 T_2} .
\end{equation}
In general, equation (\ref{totalchi}) may be expanded in powers of the
electric field squared.  Of particular interest is the Kerr
susceptibility, also in Ref.~\onlinecite{Boyd:92}:
\begin{equation}
\label{chi3}
\chi^{(3)} = \frac{4}{3} N \mu^4
\frac{T_1 T_2^2 (\Delta T_2-i)}{\hbar^3 (1+\Delta^2 T_2^2)^2} ,
\end{equation}
where $\Delta \equiv \omega-\omega_{ba}$ is the detuning of the
incoming wave from the electronic resonance frequency.  For large
detunings $\Delta T_2 \gg 1$, one obtains the approximation that:
\begin{equation}
\label{chi3-approx}
\mbox{Re}\,\chi^{(3)} \approx \frac{4}{3} N \mu^4 \left(\frac{1}{\hbar \Delta}\right)^3 \frac{T_1}{T_2}.
\end{equation}

Of course, there are many types of materials to which a simple model
of noninteracting two-level systems does not apply.  However, it has
been shown that some semiconductors such as InSb (a III-V direct bandgap
material) can be treated as a collection of independent two-level
systems with energies given by the conduction and valence bands, and
yield reasonable agreement with experiment~\cite{Miller:80}.  If the
parameter $\Delta$ is defined in terms of the bandgap energy such that
$\Delta\equiv\omega_G-\omega$, then one can look at the regime $\Delta
T_2 \gg 1$ studied above, and obtain the following equation:
\begin{equation}
\label{chi3-semiconductor}
\mbox{Re}\,\chi^{(3)} \approx -\frac{1}{30\pi \hbar^3} \left(\frac{eP}{\hbar
    \omega}\right)^4 \left( \frac{2m_r}{\hbar
    \Delta}\right)^{3/2} \frac{T_1}{T_2},
\end{equation}
where $P$ is a matrix element discussed in
Ref.~\onlinecite{Miller:80}, $\omega_G$ is the direct bandgap energy
of the system, and $m_r$ is the reduced effective mass of the exciton.
This equation displays the same scaling with lifetimes as
Eq. (\ref{chi3-approx}), so the considerations that follow should also
apply for such semiconductors.

Now, consider the effects of changing the spontaneous emission
properties for systems modeled by Eqs. (\ref{chi3-approx}) or
(\ref{chi3-semiconductor}).  When spontaneous emission is suppressed,
as in the photonic bandgap of a photonic crystal, $T_1$ will become
large while $T_2$ remains finite, thus enhancing $\chi^{(3)}$ by up to
one or more orders of magnitude (for materials with the correct
properties).  For large detunings (where $\Delta T_{2,\mbox{\tiny
    vac}} \gg 1$), we expect that $\chi^{(3)}$ will scale as $T_1 /
T_2$.  The enhancement of the real part of $\chi^{(3)}$ is defined to
be $\eta \equiv \mbox{Re}\,\chi^{(3)}_{\mbox{\tiny purcell}}\,/\,
\mbox{Re}\,\chi^{(3)}_{\mbox{\tiny hom}}$, where
$\chi^{(3)}_{\mbox{\tiny purcell}}$ is the nonlinear susceptibility in
the presence of the Purcell effect, while $\chi^{(3)}_{\mbox{\tiny
    hom}}$ is the nonlinear susceptibility in a homogeneous medium.
Since $T_1^{-1}=\Gamma_{\mbox{\tiny rad}}+\Gamma_{\mbox{\tiny nr}}$,
the maximum enhancement is predicted to be roughly:
\begin{equation}
\label{chi3enhancement}
\eta \approx \frac{T_{1,\mbox{\tiny purcell}} T_{2,\mbox{\tiny
      vac}}}{T_{1,\mbox{\tiny hom}} T_{2,\mbox{\tiny purcell}}} =
\frac{\frac{1}{2}\Gamma_{\mbox{\tiny nr}}+\gamma_{\mbox{\tiny
      phase}}}{\frac{1}{2}(\Gamma_{\mbox{\tiny nr}}+\Gamma_{\mbox{\tiny
      rad}})+\gamma_{\mbox{\tiny phase}}} \frac{\Gamma_{\mbox{\tiny
      rad}}+\Gamma_{\mbox{\tiny nr}}}{\Gamma_{\mbox{\tiny nr}}}
\end{equation}
where $\Gamma_{\mbox{\tiny rad}}$ is the radiative decay rate in
vacuum.  Since the Purcell effect increases the amplitude of
$\chi^{(1)}$, one might also expect it to increase the amplitude of
$\chi^{(3)}$; however, according to Eq.~(\ref{chi3enhancement}), the
opposite is true.  This can be understood by noting that Purcell
enhancement decreases the allowed virtual lifetime, and thus, the
likelihood of nonlinear processes to occur~\cite{Sakurai:94}.
Moreover, since the Purcell factor~\cite{Purcell:46} is calculated by
only considering the photon modes~\cite{Ryu:03}, one would not
necessarily expect phase damping effects to also play a role.
However, the results of Eq.~(\ref{chi3enhancement}) show the contrary
to be true, and can be explained as follows: when phase damping is
large, the polarization will decay quickly, thus giving rise to a
small average polarization.  However, as phase damping is lessened,
polarization decay slows down and allows the average polarization to
rise.  In the limit that phase damping is controlled exclusively by
the spontaneous emission rate, the two competing effects will cancel,
and the nonlinearity will revert to its normal value.

Furthermore, the presence of large phase damping effects makes $T_2$
effectively constant, which means that suppression of spontaneous
emission (caused by the absence of photonic states at appropriate
energies~\cite{Kleppner:81}) can enhance Kerr nonlinearities by one or
more orders of magnitude, while enhancement of spontaneous emission
via the Purcell effect~\cite{Ryu:03,Englund:05} can suppress these
nonlinearities.  For the case where Purcell enhancement takes place,
$T_1$ decreases while $T_2$ may not change as rapidly, due to the
constant contribution of phase damping effects.  This applies in the
regime where $T_1 \gg \gamma_{\mbox{\tiny phase}}^{-1}$.  Otherwise,
for sufficiently small $T_1$, $T_2$ will scale in the same way and
$\chi^{(3)}$ will remain approximately constant for large detunings,
where $\Delta T_2 \gg 1$.  This opens up the possibility of
suppressing nonlinearities in photonic crystals (to a certain degree).
For processes such as four-wave mixing or cross phase modulation,
$\chi^{(3)}$ will generally involve a detuning term and will differ
from Eq. (\ref{chi3}).  It is also interesting to note that this
enhancement scheme will generally not increase non-linear losses,
which are a very important consideration in all-optical signal
processing.  If the nonlinear switching figure of merit $\xi$ is
defined by $\xi=\mbox{Re}\,\chi^{(3)}/(\lambda \,
\mbox{Im}\,\chi^{(3)})$~\cite{Lenz:00}, then $\xi_{\mbox{\tiny
    purcell}}/\xi_{\mbox{\tiny vacuum}}=T_{2,\mbox{\tiny
    purcell}}/T_{2,\mbox{\tiny vacuum}} \geq 1$, for all cases of
suppressed spontaneous emission.

\begin{figure}[!h]
\centerline{\includegraphics[width=0.43\textwidth]{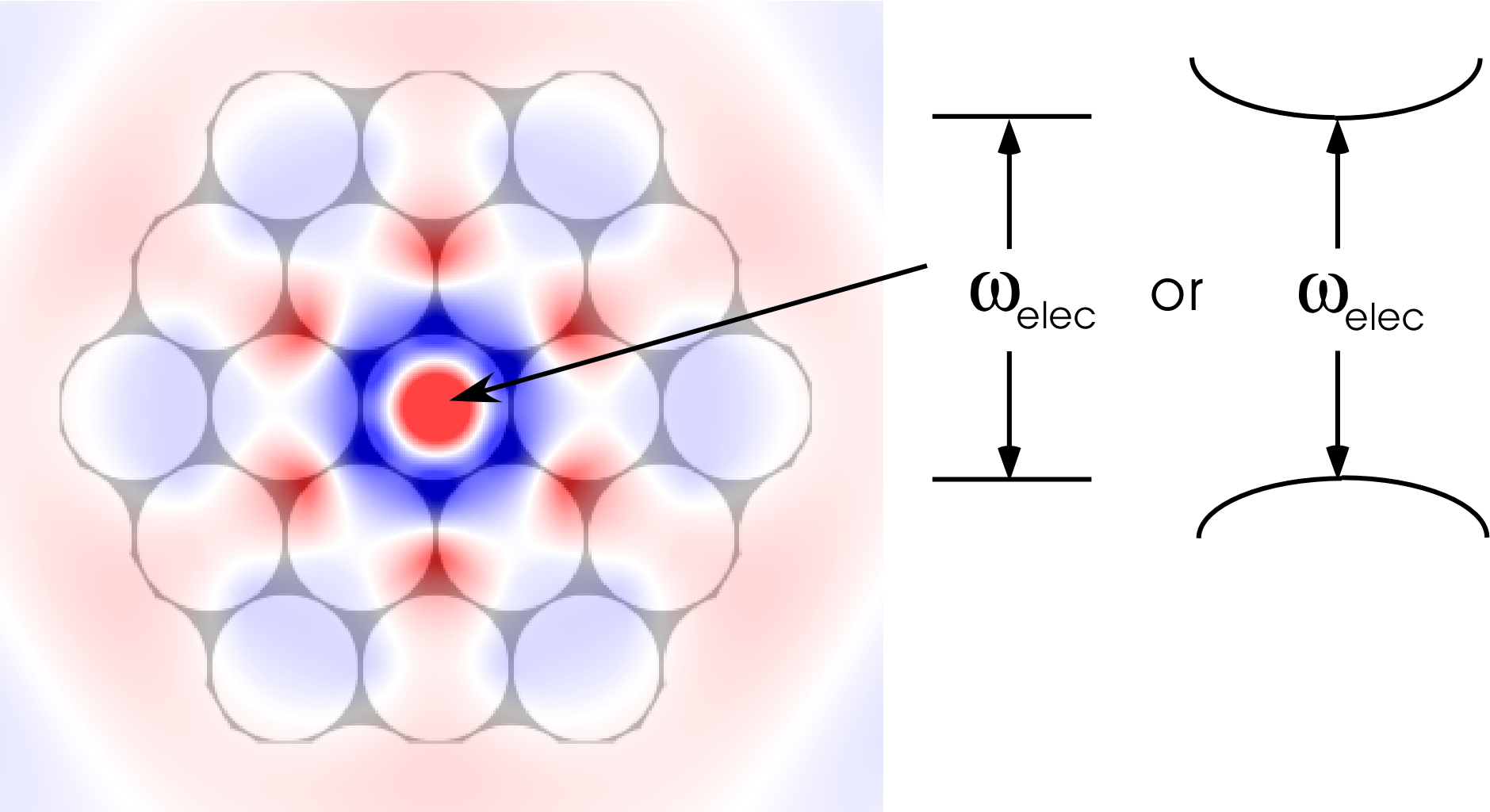}}
\caption{\label{triangularlattice} (Color) A 2D triangular lattice of air
  holes in dielectric (\( \epsilon=13\)).  On top of the dielectric
  structure in grey, the $E_z$ field is plotted, with positive values
  in red, and negative values in blue. A small region of nonlinear
  material is placed exactly in the center of the structure.  This
  material may be, for example, either two-level atoms, quantum wells,
  or some semiconductors such as InSb. }
\end{figure}

The general principal described thus far should apply for any medium
where the local density of states is substantially modified.  In what
follows, we show how this effect would manifest itself in one such
exemplary system: a photonic crystal.  This example serves as an
illustration as to how strong nonlinearity suppression / enhancement
effects could be achieved in realistic physical systems.
It consists of a 2D triangular lattice of air holes in dielectric
($\epsilon=13$), with a two-level system placed in the middle, as in
Fig.~\ref{triangularlattice}.  

Note that the vast majority of photonic crystal literature is
generally focused on modification of dispersion relations at the
frequency of the light that is sent in as a probe.  By contrast, in
the current work, it is only essential to modify the dispersion
relation for the frequencies close to the atomic resonances; the
dispersion at the frequency of the light sent in as a probe can remain
quite ordinary.

First, consider the magnitude of the enhancement or suppression of
spontaneous emission in this system.  Clearly, since there are several
periods of high contrast dielectric, two effects are to be expected.
First, there will be a substantial but incomplete suppression of
emission inside the bandgap.  Second, there will be an enhancement of
spontaneous emission outside the bandgap (since the density of states
is shifted to the frequencies surrounding the bandgap).  For an atom
polarized in the direction out of the 2-D plane, only the TM
polarization need be considered.  

\begin{figure}[!ht]
\centerline{\includegraphics[width=0.47\textwidth]{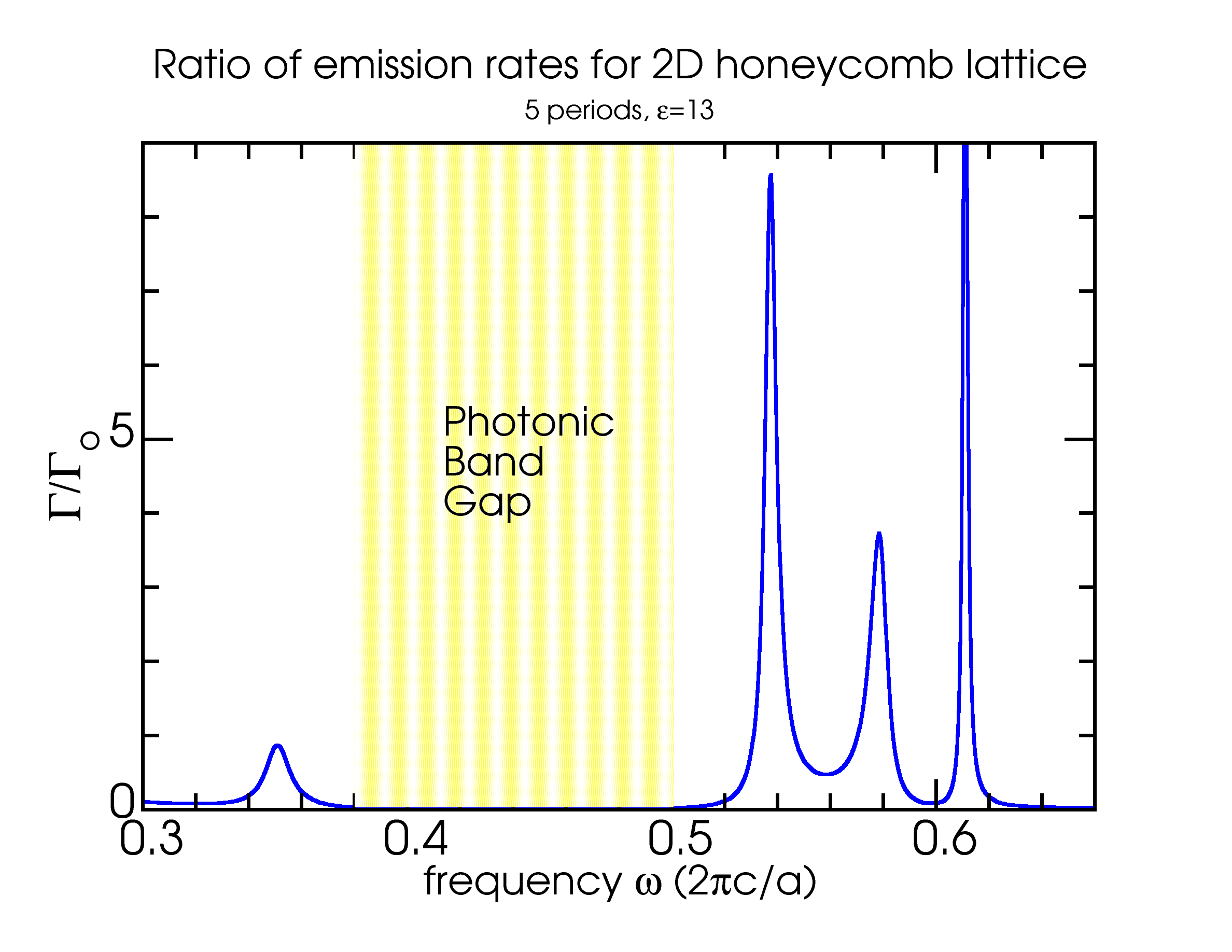}}
\caption{\label{2dphc-5periods-w0p3-0p66-q} (Color) Relative enhancement of
  the TM local density of states for Fig.~\ref{triangularlattice}, as
  measured in the time-domain simulation rate of emission, $\Gamma$,
  normalized by the emission rate in vacuum, $\Gamma_o$}
\end{figure}

\begin{figure}[!ht]
\begin{center}
\includegraphics[width=0.47\textwidth]{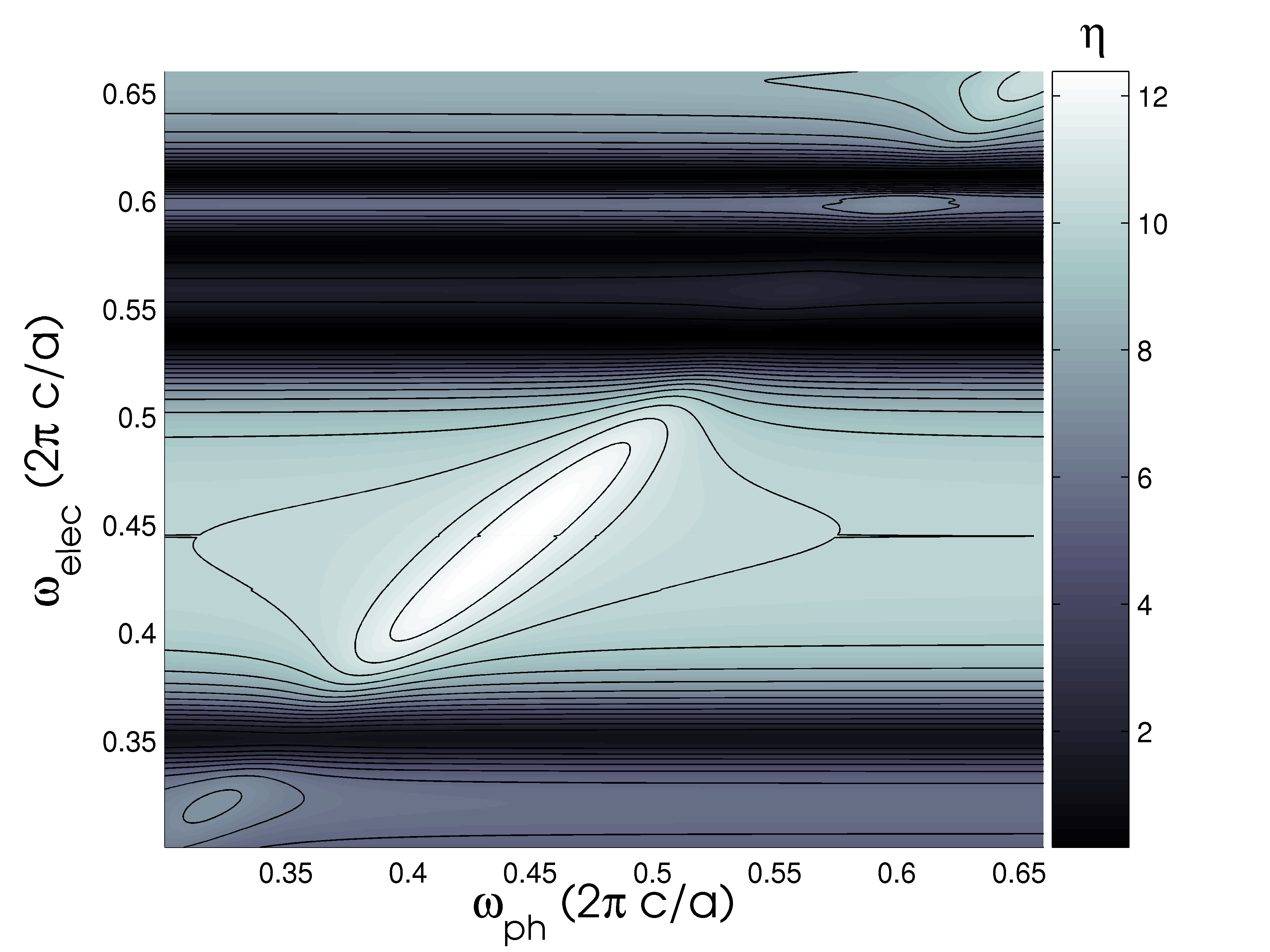}
\includegraphics[width=0.47\textwidth]{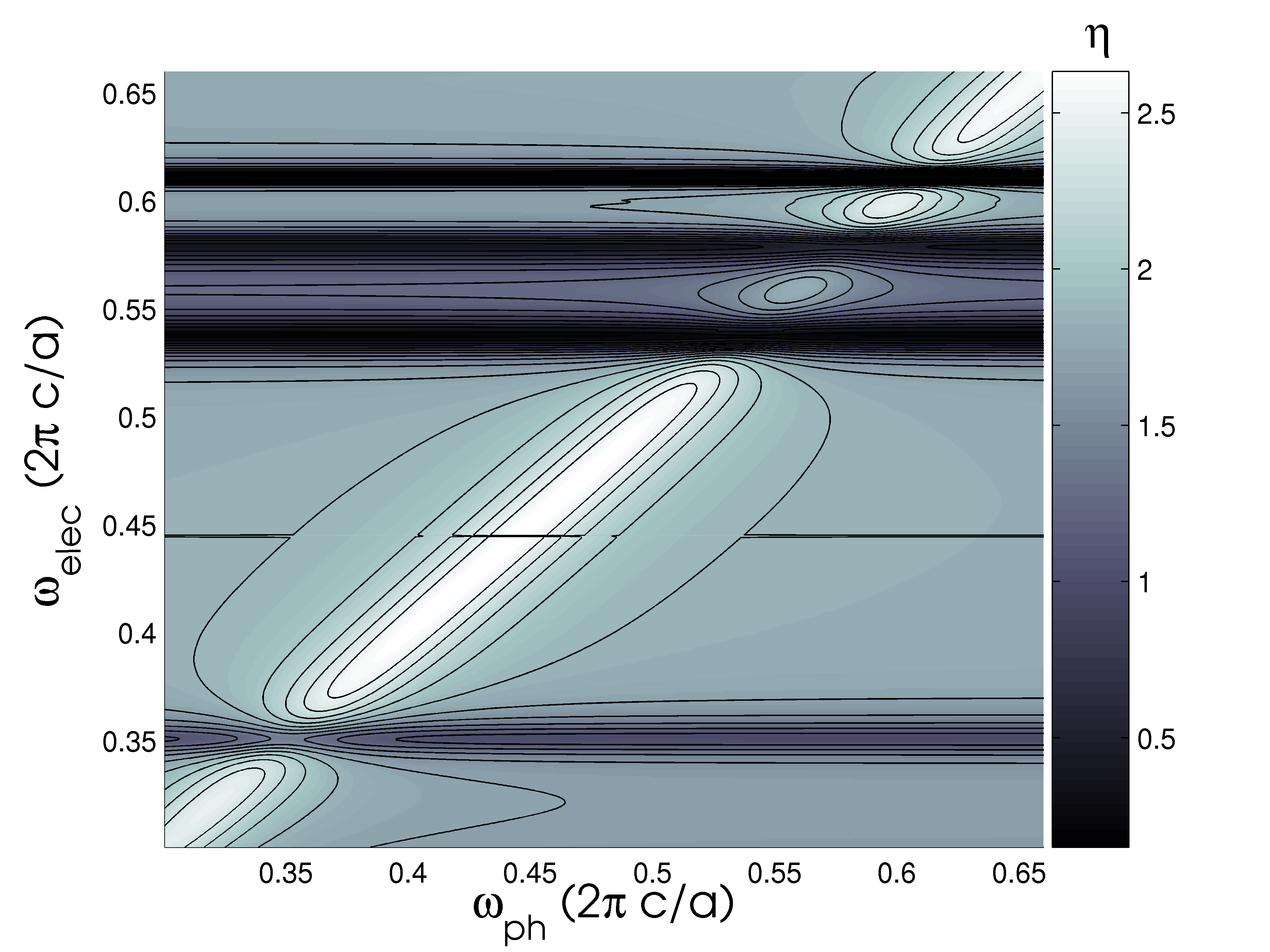}
\caption{ (Color) Contour plot of Kerr enhancement $\eta\equiv
  \mbox{Re}\,\chi^{(3)}_{\mbox{\tiny purcell}}/$
  $\mbox{Re}\,\chi^{(3)}_{\mbox{\tiny hom}}$ as a function of probe
  (\(\omega_{\mbox{\tiny ph}}\)) and electronic transition
  (\(\omega_{\mbox{\tiny elec}}\)) frequencies, for a single quantum
  well of GaAs-AlGaAs, (a) at $T=200$ K, with \(0.1\gamma_{\mbox{\tiny
      phase}}=10\Gamma_{\mbox{\tiny nr}}=\Gamma_{\mbox{\tiny rad}}\),
  and (b) at $T=225$ K, with \(0.1\gamma_{\mbox{\tiny
      phase}}=\Gamma_{\mbox{\tiny nr}}=\Gamma_{\mbox{\tiny rad}}\). }
\label{chi3-5per}
\end{center}
\end{figure}

We numerically obtain the enhancement of spontaneous emission by
performing two time-domain simulations in Meep~\cite{Farjadpour:06}, a
finite difference time-domain code which solves Maxwell's equations
exactly with no approximations, apart from discretization (which can
be systematically reduced)~\cite{Taflove:00}.  First, we calculate the
spontaneous emission of a dipole placed in the middle of the photonic
crystal structure illustrated in Fig.~\ref{triangularlattice}, then
divide by the spontaneous emission rate observed in vacuum.  The
resulting values of $T_1$ and $T_2$ are calculated numerically, and
Eq.  (\ref{chi3}), in conjunction with the definition of the
enhancement factor $\eta$, is used to plot Fig.~\ref{chi3-5per}. The
results are plotted in Fig.~\ref{2dphc-5periods-w0p3-0p66-q}.

A GaAs-AlGaAs single quantum well can lie in the interesting regime
discussed above, where the radiative loss rate $\Gamma_{\mbox{\tiny
    rad}}$ dominates the non-radiative loss rate $\Gamma_{\mbox{\tiny
    nr}}$ as well as the overall loss rate of the quantum well, for
certain temperatures~\cite{Kraiem:01}.  Equation
(\ref{chi3-semiconductor}) implies that one can see substantial
enhancement of the Kerr coefficient in that regime.

At a temperature of about 200 K, \(\Gamma_{\mbox{\tiny
    nr}}\approx0.1\Gamma_{\mbox{\tiny rad}}\)~\cite{Kraiem:01}.
Although experimental measurements for $\gamma_{\mbox{\tiny phase}}$
are unavailable to the authors, the presence of a substantial phonon
bath at that temperature leads one to expect a fairly large value,
which may be conservatively estimated by $10\Gamma_{\mbox{\tiny
    rad}}$.  These results are displayed in
Fig.~\ref{chi3-5per}(a).  Note that enhancement is
primarily observed inside the photonic bandgap (cf.
Fig.~\ref{2dphc-5periods-w0p3-0p66-q}).  We observe an enhancement in
the real part of the Kerr coefficient up to a factor of 12, close to
the predicted maximum enhancement factor of 10.48 in the regime of
large detunings ($\Delta T_2 \gg 1$).

Also, at a temperature of about 225 K, \(\Gamma_{\mbox{\tiny
    nr}}\approx\Gamma_{\mbox{\tiny rad}}\)~\cite{Kraiem:01}, and again
we take $\gamma_{\mbox{\tiny phase}}=10\Gamma_{\mbox{\tiny rad}}$.
These results are displayed in Fig.~\ref{chi3-5per}(b).  In
this case, we observe an enhancement up to a factor of 2.5, close to
the predicted maximum enhancement factor of 1.91 in the regime of
large detunings ($\Delta T_2 \gg 1$).

Finally, we note that close to room temperature (285 K), the system in
Ref.~\onlinecite{Kraiem:01} displays \(\Gamma_{\mbox{\tiny
    nr}}\approx10\Gamma_{\mbox{\tiny rad}}\), which is predicted to
yield a maximum enhancement factor of 1.06.  Since this number is
fairly negligible, it illustrates that this approach has little impact
when non-radiative losses dominate the decay of the electronic system.

On the other hand, some recent work has demonstrated that a single
quantum dot can demonstrate predominantly radiative decay in vacuum
even at room temperature, e.g., single CdSe/ZnS core-shell
nanocrystals with a peak emission wavelength of 560 nm, with
$\Gamma_{\mbox{\tiny rad}}\approx39\Gamma_{\mbox{\tiny
    nr}}$~\cite{Brokmann:04}.  Even bulk samples of similar
nanocrystals have been shown to yield a significant radiative decay
component, corresponding to $\Gamma_{\mbox{\tiny rad}} \approx
\Gamma_{\mbox{\tiny nr}}$~\cite{Talapin:01}.  Thus, we predict that with
strong suppression of radiative decay, nonlinear enhancement of a factor 
of two, or more, could be observed at room temperature.

We now discuss the implications of this effect on previous work
describing nonlinearities in photonic crystals, such as
Ref.~\onlinecite{Soljacic:04}, and the references therein.  Most past
experiments should not have observed this effect, because they were
designed with photonic bandgaps at optical frequencies significantly
smaller than the frequencies of the electronic resonances generating
the nonlinearities, in order to operate in a low-loss regime.
Furthermore, in most materials, non-radiative decays will dominate
radiative decays at room temperature.  Finally, all the previous
\emph{analyses} are still valid as long as one considers the input
parameters to be effective nonlinear susceptibilities, which come from
natural nonlinear susceptibilities modified in the way described by
this paper.

In conclusion, we have shown that the Purcell effect can be used
to tailor optical nonlinearities. This principle manifests itself in
an exemplary two-level system embedded in a photonic crystal; for
realistic physical parameters, enhancement of Kerr nonlinearities by
more than an order of magnitude is predicted. The described phenomena
is caused by modifications of the local density of states near the
resonant frequency.  Thus, this treatment can easily be applied to
analyze the Kerr nonlinearities of two-level systems in almost any
geometrical structure in which the Purcell effect is substantial
(e.g., photonic crystal fibers~\cite{Litchinitser:02}, optical
cavities).  It also presents a reliable model for a variety of
materials, such as quantum dots, atoms, and certain semiconductors.
Future investigations will involve extending the formalism in this
manuscript to other material systems.

The authors would like to thank Robert Boyd, Erich Ippen, Daniel
Kleppner, Vladan Vuletic, Moti Segev, Steven G.  Johnson, and Mark
Rudner for valuable discussions.  This work was supported in part by
the Army Research Office through the Institute for Soldier
Nanotechnologies under Contract No. DAAD-19-02-D0002.  A.R.
acknowledges the support of the Department of Energy under Grant No.
DE-FG02-97ER25308.


\begin{thebibliography}{10}

\bibitem{Drazin:89}
P. Drazin and R. Johnson, {\em Solitons: an Introduction} (Cambridge University
  Press, Cambridge, England, 1989).

\bibitem{Boyd:92}
R.~W. Boyd, {\em Nonlinear Optics} (Academic Press, San Diego, 1992).

\bibitem{Nielsen:00}
M. Nielsen and I. Chuang, {\em Quantum Computation and Quantum Information}
  (Cambridge University Press, Cambridge, England, 2000).

\bibitem{Agrawal:01}
G. Agrawal, {\em Applications of Nonlinear Fiber Optics,}, {\em Optics and
  Photonics} (Academic Press, San Diego, CA, 2001).

\bibitem{Purcell:46}
E. Purcell, Phys. Rev. {\bf 69},  681  (1946).

\bibitem{Kleppner:81}
D. Kleppner, Phys. Rev. Lett. {\bf 47},  233  (1981).

\bibitem{Ryu:03}
H. Ryu and M. Notomi, Opt. Lett. {\bf 28},  2390  (2003).

\bibitem{Bermel:04}
P. Bermel, J.~D. Joannopoulos, Y. Fink, P.~A. Lane, and C. Tapalian, Phys. 
  Rev. B {\bf 69},  035316  (2004).

\bibitem{Englund:05}
D. Englund, D. Fattal, E. Waks, G. Solomon, B. Zhang, T. Nakaoka, Y. Arakawa,
  Y. Yamamoto, and J. Vuckovic, Phys. Rev. Lett. {\bf 95},  013904  (2005).

\bibitem{JohnQu:96}
S. John and T. Quang, Phys. Rev. Lett. {\bf 76},  2484  (1996).

\bibitem{Preskill:04}
J. Preskill, Quantum Computation Lecture Notes,
  http://www.theory.caltech.edu/people/preskill/ph229/, 2004.

\bibitem{CohenTannoudji:77}
C. Cohen-Tannoudji, B. Diu, and F. Lalo\"{e}, {\em Quantum Mechanics} (John
  Wiley and Sons, New York, 1977).

\bibitem{John:94}
S. John and T. Quang, Phys. Rev. A {\bf 50},  1764  (1994).

\bibitem{Lambropoulos:00}
P. Lambropoulos, G.~M. Nikolopoulos, T.~R. Nielsen, and S. Bay, Rep. Prog.
  Phys. {\bf 63},  455  (2000).

\bibitem{Bayer:01}
M. Bayer, T.~L. Reinecke, F. Weidner, A. Larionov, A. McDonald, and A. Forchel,
  Phys. Rev. Lett. {\bf 86},  3168  (2001).

\bibitem{Lodahl:04}
P. Lodahl, A.~F. van Driel, I.~S. Nikolaev, A. Irman, K. Overgaag, D.
  Vanmaekelbergh, and W.~L. Vos, Nature {\bf 430},  654  (2004).

\bibitem{Miller:80}
D. Miller, S. Smith, and B. Wherrett, Opt. Comm. {\bf 35},  221  (1980).

\bibitem{Sakurai:94}
J.~J. Sakurai, {\em Modern Quantum Mechanics} (Addison-Wesley, Reading, MA,
  1994).

\bibitem{Lenz:00}
G. Lenz, J. Zimmermann, T. Katsufuji, M.~E. Lines, H.~Y. Hwang, S. Spälter,
  R.~E. Slusher, S.-W. Cheong, J.~S. Sanghera, and I.~D. Aggarwal, Opt. Lett.
  {\bf 25},  254  (2000).

\bibitem{Farjadpour:06}
A. Farjadpour, D. Roundy, A. Rodriguez, M. Ibanescu, P. Bermel, J.~D.
  Joannopoulos, S.~G. Johnson, and G. Burr, Opt. Lett. {\bf 31},  2972  (2006).

\bibitem{Taflove:00}
A. Taflove and S.~C. Hagness, {\em Computational Electrodynamics}, 2nd ed.
  (Artech House, Norwood, MA, 2000).

\bibitem{Kraiem:01}
S. Kraiem, F. Hassen, H. Maaref, X. Marie, and E. Vaneelle, Opt. Mat. {\bf 17},
   305  (2001).

\bibitem{Brokmann:04}
X. Brokmann, L. Coolen, M. Dahan, and J.~P. Hermier,
Phys. Rev. Lett. {\bf 93}, 107403 (2004).

\bibitem{Talapin:01}
D.~V. Talapin, A.~L. Rogach, A. Kornowski, M. Haase, and H. Weller, Nano Lett. {\bf 1}, 207 (2001).

\bibitem{Soljacic:04}
M. Soljacic and J.~D. Joannopoulos, Nature Mat. {\bf 3}, 211 (2004).

\bibitem{Litchinitser:02}
N.~M. Litchinitser, A. Abeeluck, C. Headley, and B. Eggleton, Opt. Lett. {\bf
  27},  1592  (2002).

\end{thebibliography}
\end{document}